\documentclass[%
 reprint,
 amsmath,amssymb,
 aps,
 pra,
]{revtex4-2}

\usepackage{graphicx}
\usepackage{dcolumn}
\usepackage{bm}
\usepackage{amsmath}
\usepackage{comment}
\usepackage[dvipsnames]{xcolor}
\usepackage{subfigure}
\DeclareMathOperator*{\SumInt}{%
\mathchoice%
  {\ooalign{$\displaystyle\sum$\cr\hidewidth$\displaystyle\int$\hidewidth\cr}}
  {\ooalign{\raisebox{.14\height}{\scalebox{.7}{$\textstyle\sum$}}\cr\hidewidth$\textstyle\int$\hidewidth\cr}}
  {\ooalign{\raisebox{.2\height}{\scalebox{.6}{$\scriptstyle\sum$}}\cr$\scriptstyle\int$\cr}}
  {\ooalign{\raisebox{.2\height}{\scalebox{.6}{$\scriptstyle\sum$}}\cr$\scriptstyle\int$\cr}}
}

\begin{document}

\preprint{APS/123-QED}

\title{Resonant control of photoelectron directionality by \\ interfering one- and two-photon pathways}

\author{Yimeng Wang}
\author{Chris H. Greene}%
\affiliation{%
 Department of Physics and Astronomy, Purdue University, West Lafayette, Indiana 47907, USA and
Purdue Quantum Science and Engineering Institute,
Purdue University, West Lafayette, Indiana 47907, USA
}%

\date{\today}

\begin{abstract}
Coherent control of interfering one- and two-photon processes has for decades been the subject of research to achieve the redirection of photocurrent. The present study develops two-pathway coherent control of ground-state helium atom above-threshold photoionization for energies up to the $N=2$ threshold, based on a multichannel quantum defect and $R$-matrix calculation. 
Three parameters are controlled in our treatment: the optical interference phase $\Delta\Phi$, the reduced electric field strength $\chi=\mathcal{E}_{\omega}^2/{\mathcal{E}_{2\omega}}$ , and the final state energy $\epsilon$. A small energy change near a resonance is shown to flip the emission direction of photoelectrons with high efficiency, through an example where $90\%$ of photoelectrons whose energy is near the $2p^2\ ^1S^e$ resonance flip their emission direction. However, the large fraction of photoelectrons ionized at the intermediate state energy, which are not influenced by the optical control, make this control scheme challenging to realize experimentally.
\end{abstract}

\maketitle
\section{Introduction}
Coherent control scenarios have generated extensive attention in condensed matter systems and in atomic and molecular physics. The basic idea is to introduce a difference in two alternative electric dipole transition amplitudes in order to manipulate the interference between them and thereby control an observable outcome. 
In particular, phase-sensitive coherently controlled quantum interference enables observation of novel physics. 
For example, the two-color  phase-sensitive coherent control can be applied to various scenarios in physical chemistry and molecular physics in order to control the branching ratio among different reaction products \cite{Zhu77,PhysRevLett.92.113002,PhysRevLett.74.4799}, to rotate the molecular polarization, and to selectively ionize oriented molecules \cite{PhysRevLett.96.173001}. In condensed matter physics it is primarily of interest to control the current flow direction in a semiconductor \cite{PhysRevLett.74.3596,PhysRevLett.92.147403}, and in quantum computation to depress the linkage error of qubits \cite{PhysRevLett.111.233002}.  This technique is also used in femtosecond and attosecond experiments \cite{PhysRevLett.117.217601} and in the strong field regime \cite{PhysRevLett.73.1344}, and to achieve quantum path control between short and long electron trajectories \cite{PhysRevLett.107.153902}.


Compared with the extensive experimental literature, there are comparatively few theoretical calculations that provide a full treatment of such coherently controlled systems \cite{Anderson:1992}. Several calculations have been carried out for photoionization of Ne \cite{PhysRevA.100.063417,PhysRevX.10.031070}, H$_2$ \cite{PhysRevLett.86.5454}, and dc-field dressed hydrogen and alkali-metal atoms in a limited energy range \cite{PhysRevLett.82.4791,PhysRevA.100.043409}.   
The present study treats the $\omega-2\omega$ coherent control of helium ionization, an atom for which the electron correlations have been extensively calculated and interpreted \cite{Ho:1983,Ho:1985,PhysRevA.39.115,PhysRevLett.78.4902,Petersen1991,MeyerGreene1994pra}. 
The present study computes the photoelectron angular distribution (PAD) to analyze the phase dependence of the directional right-left (or upper-lower) asymmetry parameter, especially as influenced by Fano-Feshbach resonances, and the role of autoionizing states in affecting the interference between one- and two-photon ionization pathways. 
In contrast to previous studies of the coherent control of photoelectron branching ratios into multiple open channels \cite{PhysRevLett.79.4108,PhysRevLett.82.65,PhysRevA.76.053401,PhysRevA.80.033401}, the present treatment considers ionization into a single open channel that possesses, however, three contributing partial waves.

\section{Theory}

\begin{figure}[htbp]
  \includegraphics[scale=0.45]{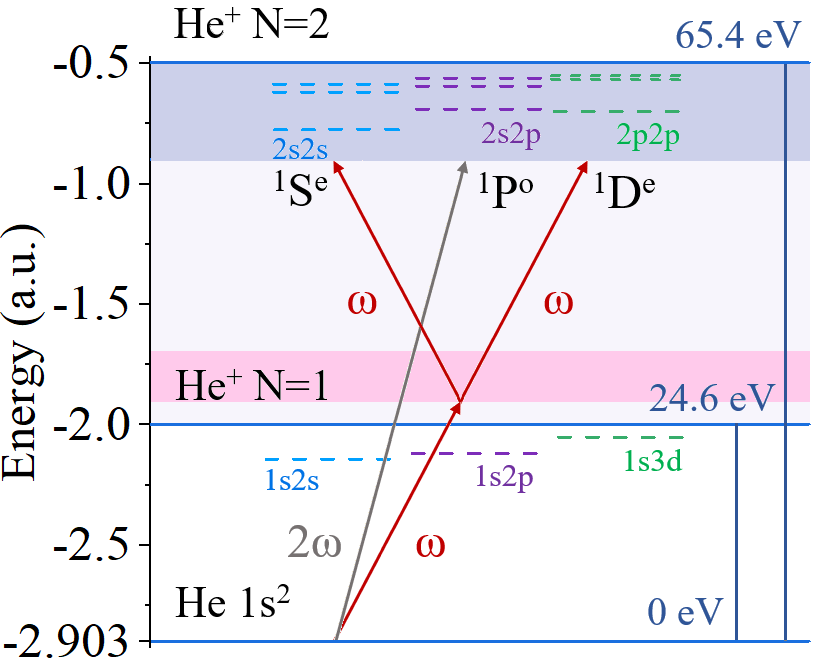}
    \caption{ Schematic energy level diagram of helium for the relevant transitions. The He ground state is ionized, with the one and two-photoionization pathways indicated by arrows. The shaded regions show the energy ranges considered for the final and intermediate states, both of which are between the $N=1$ and $N=2$ thresholds. The dashed lines give the lowest few bound and autoionizing energy levels\cite{Ho:1983} below the $N=1$ and $N=2$ thresholds, respectively, with $^{2S+1}L^{\pi}$ spectroscopic labels from left to right as $^{1}S^{e}$,$^{1}P^{o}$ and $^{1}D^{e}$. The autoionization levels that are relevant to our calculations are those above the $N=1$ thresholds. There are no intermediate-state resonances involved in the frequency range considered, only an open ionization continuum. }
  \label{1}
\end{figure}

The bichromatic laser electric field considered in our treatment is given by $\Vec{\mathcal{E}}(t)$: 

\begin{equation}
    \Vec{\mathcal{E}}(t)= \hat{\epsilon}\left(\mathcal{E}_{2\omega}e^{-i(2\omega t+\Phi_{2\omega})}+\mathcal{E}_{\omega}e^{-i(\omega t+\Phi_{\omega})} +c.c.\right).
\end{equation}
Here $\mathcal{E}_{\omega,2\omega}$ are the electric field amplitudes for the fundamental and second harmonic. 
The two fields have a variable but well-defined phase relation, denoted by $\Phi_{\omega,2\omega}$, and both of the fields are chosen here to be linear-polarized along a common $z$-axis {\it i.e.}, $\hat{\epsilon}=\hat{z}$. The frequency range considered is $\omega=1.0-1.2$ $a.u.$. The schematic diagram of the ionization process is shown in Fig.~\ref{1}. A ground state He atom at $E_g=-2.90$ $a.u.$ absorbs either one photon with energy $2\omega$ or two photons with each energy $\omega$, reaching a final state $f$ with energy from $-0.9$ to $-0.5$ $a.u.$ (indicated by the upper shaded region of Fig. \ref{1}). 
The two-photon pathway is an above threshold ionization (ATI), with intermediate energies given by the lower shaded region of Fig.\ref{1}. The resonances converging to the $N=2$ threshold are of particular interest. 
The atomic orbital angular momentum is initially $L_i=0$, and it changes after absorption of one electric dipole photon to $L_f=1$, or after two-photon absorption to $L_f=0,2$. The parity $\pi$ flips between even and odd for each photon absorption step, and the atomic spin $S$ remains in the singlet state since spin-spin and spin-orbit interactions are neglected in this study. 

It is well known that the $\omega-2\omega$ scheme displays no interference effects that can influence the total yield \cite{PhysRevLett.69.2353}. This is because the even- and odd-parity final states are in principle distinguishable, which implies that no interference occurs in any observable that commutes with the parity operator such as the integrated absorption rate. However, the $\omega-2\omega$ scheme influences the angular emission of the photoelectron, since an angular observable represents an operator that does not commute with parity.  This interference has been confirmed by  experiments \cite{PhysRevLett.69.2353,PhysRevA.76.053401,PhysRevA.80.033401,PhysRevLett.98.053001}, which show that tuning the phase difference $\Delta\Phi=2\Phi_{\omega}-\Phi_{2\omega}$ causes a sinusoidal modulation that can be observed in the integrated lower- or upper- (negative or positive $z$) dominance in the photoejection directions. The remainder of this article shows the sinusoidal modulation in the computed photoelectron angular distribution $\frac{dW(\theta)}{d\Omega}$: 


\begin{equation}
\label{eq-1}
\begin{split}
    \frac{dW(\theta)}{d\Omega}&=\left|c_0Y_{00}(\theta)e^{i\phi_0}+c_1Y_{10}(\theta)e^{i(\phi_1+\Delta\Phi)}+c_2Y_{20}(\theta)e^{i\phi_2}\right|^2\\
    &= \frac{W_{\rm tot}}{4\pi} \sum_{j=0}^{4}\beta_j P_j(\cos{\theta})
\end{split}
\end{equation}
Here $\theta$ is the polar angle between the ejected electron and the polarization axis; there is no $\phi$ dependence owing to the azimuthal symmetry.
$W_{\rm tot}$ is the angle-integrated transition rate and $\beta_0 \equiv 1$.
In the first line of Eq. \ref{eq-1}, the differential transition rate $\frac{dW(\theta)}{d\Omega}$ is given by a coherent sum of the different angular components, with complex transition amplitude $c_{l}e^{i\phi_{l}}$ for partial wave $l$, where $c_{l}$ is real and positive. Since a photoelectron in our calculation can escape only with He$^+$ in the $1s$ state, its angular momentum takes the values $l=L_f=0,1,2$. The experimentally-controllable optical phase is $\Delta\Phi=2\Phi_{\omega}-\Phi_{2\omega}$,  which is distinct from  the intrinsic phases in the amplitudes $\phi_{l}$ that reflect the atomic physics.  Note that the latter are strongly energy dependent near resonances and thresholds: they include contributions from the long-range Coulomb potential, the electron correlations, and the intermediate scattering states. 

The second line of Eq. \ref{eq-1} rearranges the summed products of spherical harmonics $Y_{l0}(\theta)$ into Legendre polynomials $P_j(\cos{\theta})$ with real coefficients $\beta_j$. The even (odd) order $P_{j}(\cos{\theta})$ gives the symmetric (anti-symmetric) photoelectron distribution along $\theta=\pi/2$ which produce differences between  the lower- and upper halves of the emission sphere, i.e., the negative and positive $z$ regions, respectively. In the absence of interference, the odd orders of $P_{j}(\cos{\theta})$ would vanish and no asymmetry would be observed between the two hemispheres.  With some specific values of $\beta_j$, it is possible to guide most electrons to one side, as we will demonstrate in the latter discussion.  
A directional asymmetry parameter $\alpha_L=W_L/(W_L+W_U)$ has been measured in some experiments\cite{PhysRevLett.69.2353,PhysRevA.76.053401,PhysRevA.80.033401,PhysRevLett.98.053001}, so we use it to quantify the ratio between the lower-directed electron current and the total: 
\begin{equation}
\begin{split}
    W_L&= 2\pi \int_{\frac{\pi}{2}}^{\pi} \frac{dW(\theta)}{d\Omega}\sin \theta d\theta = \frac{W_{\rm tot}}{2} \left ( 1-\frac{1}{2}\beta_1+\frac{1}{8}\beta_3 \right )  \\
    W_U&= 2\pi\int_0^{\frac{\pi}{2}} \frac{dW(\theta)}{d\Omega}\sin \theta d\theta = \frac{W_{\rm tot}}{2} \left ( 1+\frac{1}{2}\beta_1-\frac{1}{8}\beta_3 \right ) \\
\end{split}    
\end{equation}

For $\alpha_L=1$ (or $0$), all the photoelectrons go to the lower (upper) side, while at $\alpha_L=0.5$, there is no preference over either direction; this usually happens at resonances when one of the definite parity pathways is dominant. $\beta_{j}$ and the total rate $W_{\rm tot}$ in terms of the transition amplitudes $c_{l}e^{i\phi_{l}}$ are given here: 
\begin{equation}
\begin{split}
W_{\rm tot}&={c_0}^2+{c_1}^2+{c_2}^2\\
W_{\rm tot}\beta_1&=2\sqrt{3} c_0c_1\cos{[\Delta\Phi-(\phi_0-\phi_1)]}\\
&+ 4\sqrt{\frac{3}{5}} c_1c_2 \cos{[\Delta\Phi-(\phi_2-\phi_1)]}\\
W_{\rm tot}\beta_2&=2{c_1}^2+\frac{10}{7}{c_2}^2+2\sqrt{5} c_0c_2 \cos{(\phi_2-\phi_0)}\\
W_{\rm tot}\beta_3&=6\sqrt{\frac{3}{5}} c_1c_2 \cos{[\Delta\Phi-(\phi_2-\phi_1)]}\\
W_{\rm tot}\beta_4&=\frac{18}{7} {c_2}^2
\end{split}    
\end{equation}
Therefore the directional asymmetry parameter $\alpha_L$ is:
\begin{equation}
\label{eq-2}
\begin{split}
    \alpha_L&=\frac{1}{2}-\frac{\sqrt{3}}{2}\frac{c_0c_1}{{c_0}^2+{c_1}^2+{c_2}^2}\cos{[\Delta\Phi-(\phi_0-\phi_1)]}\\
    &-\frac{\sqrt{15}}{8}\frac{c_1c_2}{{c_0}^2+{c_1}^2+{c_2}^2}\cos{[\Delta\Phi-(\phi_2-\phi_1)]}\\
    &\equiv\frac{1}{2}+A(\chi,\epsilon) \cos{[\Delta\Phi-\varphi(\epsilon)]}
\end{split}    
\end{equation}
The second equality of Eq. \ref{eq-2} recasts the directional asymmetry parameter $\alpha_L$ in terms of an amplitude $A(\chi,\epsilon)$ and a phase $\varphi(\epsilon)$, both of which are energy sensitive ($\epsilon=E_g+2\omega$ is the final state energy). The amplitude $A$ also depends on the electric fields $\mathcal{E}_{\omega,2\omega}$, as a function of $\chi=\mathcal{E}_{\omega}^2/{\mathcal{E}_{2\omega}}$, while $\varphi$ is independent of the field strengths.  $0\leq A\leq \frac{1}{2}$ and $0\leq\varphi\leq2\pi$. In order to maximize $\alpha_L$, $\varphi$ should equal the optical phase difference. 
As we will show, $\varphi$ is encoded with electron-correlation information as are the phases $\phi_i$.


The transition amplitudes $c_{l}e^{i\phi_{l}}$ for weak electric fields can be computed using perturbation theory \cite{Fano:1973,PhysRevA.103.033103}:  
\begin{equation}
\label{eq-4}
\begin{split}
    c_0e^{i\phi_0}&=\frac{\sqrt{2\pi}}{3}\mathcal{E}_{\omega}^2 \langle f_0|\mathcal{G}(\omega,\Vec{r}^{\prime},\Vec{r})\Vec{r}^{\prime}\cdot\Vec{r}|i\rangle\\%
    c_1e^{i\phi_1}&=\sqrt{2\pi}\mathcal{E}_{2\omega}\langle f_1|r_z|i\rangle\\
    c_2e^{i\phi_2}&=\frac{\sqrt{2\pi}}{3}\mathcal{E}_{\omega}^2 \langle f_2|\mathcal{G}(\omega,\Vec{r}^{\prime},\Vec{r})(3r_z^{\prime} r_z-\Vec{r}^{\prime}\cdot\Vec{r})|i\rangle\\
\end{split}    
\end{equation}
Here $|i\rangle$ and $|f_{l}\rangle$ are the energy eigenstates for the unperturbed helium atom. In the formula above, we rearrange the dipole operators: usually for one- and two-photon transitions we have 
$d^{(1)}=\Vec{r}\cdot\hat{\epsilon}$ and $d^{(2)}=(\Vec{r}^{\prime}\cdot\hat{\epsilon})(\Vec{r}\cdot\hat{\epsilon})$ (where vector operator $\Vec{r}=\Vec{r}_1+\Vec{r}_2$ is the sum for the two electrons in helium; dipole approximation is applied). 
In Eq. \ref{eq-4} the single-photon and two-photon electric dipole transition operators are written as rank-0,-1 and -2 tensors for different $l$. 
For the two-photon amplitudes, the Green's function is introduced for the intermediate ATI transition and can be written formally as: 
\begin{equation}
\label{eq-5}
\mathcal{G}(\omega,\Vec{r}^{\prime},\Vec{r})=\SumInt_m\frac{\langle\Vec{r}^{\prime}|m\rangle\langle m|\Vec{r}\rangle}{ \omega-\Delta_{mi}}
\end{equation}
$\Delta_{mi}=E_m-E_g$, where intermediate energies $E_m$ include all the eigenvalues of the unperturbed helium Hamiltonian that obey the parity and angular momentum selection rules for single-photon ionization \cite{Robicheaux:1993}. A mixed notation $\SumInt_m$ of summation and integration is used, because $|m\rangle$ includes both bound and continuum states with different normalizations.

Equation \ref{eq-4} are computed using generalized multi-channel quantum defect (MQDT) \cite{Greene:1979,Seaton:1983,GRF1982,ErratumGRF1982} and the streamlined $R$-matrix method \cite{Review:1996}. In our calculation, an artificial boundary is set at the radius $R_0=15$ $a.u.$ from the nucleus, within which the electron-electron interactions will be fully considered. For the region outside the boundary, the Gailitis-Damburg transformation \cite{GailitisDamburg,SGC1992,SC1993} is used to incorporate the electron correlations up to the second-order in $1/r$. 
The Green's function $\mathcal{G}(\omega,\Vec{r}^{\prime},\Vec{r})$(Eq. \ref{eq-5}) is solved through an inhomogeneous R-matrix method implemented by Robicheaux and Gao \cite{Robicheaux:1991,Robicheaux:1993}. The details of all these methods can be found in Ref.\cite{PhysRevA.103.033103}
Our calculation follows the PAD results treated in previous studies such as Refs.\cite{PhysRevA.103.033103,Diego:2019,Lindsay:1992}.

\section{Results and Discussion}

\begin{figure}[htbp]
  \includegraphics[scale=0.31]{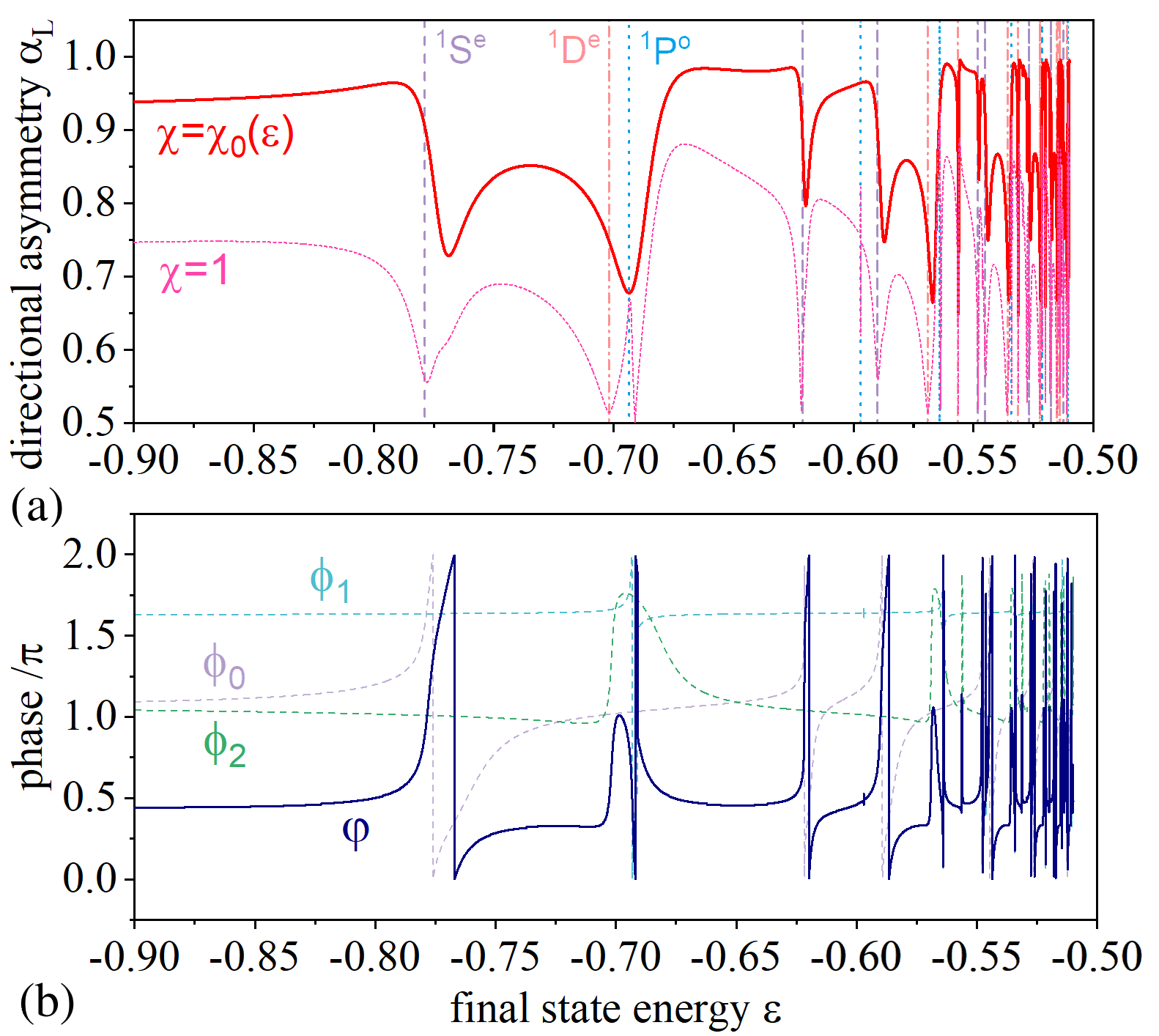}
\caption{ (a). The solid curve shows the maximized ratio of lower-oriented electrons $\alpha_L^{max}$ obtained by optimizing the reduced field strength $\chi=\mathcal{E}_{\omega}^2/{\mathcal{E}_{2\omega}}$ and optical phase $\Delta\Phi$ at each energy. The dashed curve shows the $\alpha_L$ when $\chi=1$. The background vertical lines give the position of resonances (with different symmetries indicated by different line-types); those positions are near the local minima of the $\alpha_L$-curves. 
(b). The directional asymmetry phase $\varphi$ as the solid curve, namely the optical phase corresponding to $\alpha_L^{max}$. $\varphi$ experiences a dramatic change over $2\pi$ or $\pm\pi$ across a resonance, which can be traced back to the dipole transition moment phases $\phi_{l}$ that are shown as dashed curves. }
  \label{2}
\end{figure}

In this paper, the optical control of $\alpha_L$ will be discussed in terms of three parameters: the relative laser phase $\Delta\Phi$, the reduced field strength $\chi=\mathcal{E}_{\omega}^2/{\mathcal{E}_{2\omega}}$, and the final state energy $\epsilon$ that plays a major role owing to the existence of resonances. The computation of $A(\chi,\epsilon)$ and $\varphi(\epsilon)$ versus energy is the main feature of our work, plus the identification of regions where very high control is readily achievable.  
With the knowledge of $\alpha_L$, the redirection of photoelectron emission can be discussed in a more complete manner, as control can be optimized by choosing energies that maximize the directional asymmetry. In addition, photocurrents can be redirected not only through phase control, but also by tuning the photon frequency with fixed phases and field strengths.  

First, consider our results for $\alpha_L^{max}$ when the reduced field strength $\chi$ and optical phase difference $\Delta\Phi$ are optimized at each energy. To maximize $\alpha_L$ the choice of optical phase is $\Delta\Phi=\varphi(\epsilon)$, (Fig.~\ref{2} (b)). The choice of $\chi$ is determined by writing the amplitude $A$ in the form $A(\chi,\epsilon)\propto\chi/(a(\epsilon)+b(\epsilon)\chi^2)$, with $a(\epsilon)$ and $b(\epsilon)$ being determined from Eq. \ref{eq-4}. A single peak of $A(\chi,\epsilon)$ exists at $\chi_0(\epsilon)$. Note that whatever is the value of $\chi$, there is no influence on the value of $\varphi$, and therefore the two parameters can be tuned separately and independently. By selecting a proper $\chi$, we can largely improve the efficiency of coherent phase control of the directional electron photoemission.

Our calculated $\alpha_L^{max}=1/2+A(\chi_0(\epsilon),\epsilon)$ is plotted in Fig.~\ref{2} (a), as the solid curve. 
Observe that $\alpha_L^{max}$ ranges between $0.65-1.0$ and indicates a quite high-efficiency level of control. However, at energies near the resonances, which are the regions of our greatest interest, $\alpha_L^{max}$ drops down. This is expected since the symmetry of any specific resonance has a single transition amplitude there that overwhelms the amplitude from other channels, and the PAD behaves as if only a single path-way is allowed. This is even more obvious for $\alpha_L$ at $\chi=1$ a.u. (shown as the thin dashed curve), where $\alpha_L$ drops back to $0.5$ at almost every resonance energy. Accordingly, tuning the reduced field strength $\chi$ can alleviate but not eliminate the asymmetry-diminishing tendency at resonance energies.

The directional asymmetry phase $\varphi(\epsilon)$ is presented in Fig.~\ref{2} (b). When far away from the resonance, $\varphi(\epsilon)$ experiences a small change over the whole range,  fluctuating only over $0.55\pi-0.7\pi$. However, across the resonance it changes dramatically over the full possible range of $2\pi$ or $\pm\pi$.  These features of $\varphi(\epsilon)$ would enable frequency-sensitive coherent-control to redirect the photocurrent, as is discussed below.
The changes of $\varphi(\epsilon)$ show a behavior similar to that of the dipole transition phases $\phi_l$ (the dashed curves), which are properties of the final and intermediate scattering states. The wave function of a detected photoelectron satisfies the incoming wave boundary condition \cite{PhysRev.93.888}: It approaches the outgoing wave portions of a plane wave pointing towards the detector at infinity, implying that the scattering wave function can parameterized as, $ \psi_f(r)\rightarrow\frac{1}{\sqrt{2\pi k}}\left(e^{ikr}r^{i/k}-e^{-2i\delta_l}e^{-ikr}r^{-i/k}\right)$, where $\delta_l=\eta_l+\pi\tau_l$, with $\eta_l$ the Coulomb phase parameters: $\eta_l=\frac{\ln (2 k)}{k} + arg[\Gamma(l + 1 - \frac{i}{k})] - \frac{l\pi}{2}$, and $\tau_l$ incorporate both the quantum defects and the influences from all the closed channels\cite{Greene:1985}. For the single-photon transition we have $\phi_1=\delta_1$.
For a two-photon ATI process, there is an extra phase that comes from $\mathcal{G}(\omega,\Vec{r}^{\prime},\Vec{r})$. Since the undetected intermediate scattering wave can be treated as purely outgoing at large distance \cite{Robicheaux:1993}, $\mathcal{G}(\omega,\Vec{r}^{\prime},\Vec{r})$ can be written into a principal value part and an ``on shell'' part as \cite{Greene:1979}: 
\begin{equation}
\label{eq-3}
 \mathcal{G}(\omega,\Vec{r}^{\prime},\Vec{r})=\mathcal{G}^{(\mathcal{P})}(\omega,\Vec{r}^{\prime},\Vec{r}) -i\pi \langle\Vec{r}^{\prime}|m^{(sh)}\rangle\langle m^{(sh)}|\Vec{r}\rangle
\end{equation}
where $\mathcal{G}^{(\mathcal{P})}(\omega,\Vec{r}^{\prime},\Vec{r})$ is the principal value Green's function, and the on-shell state $\langle\Vec{r}|m^{(sh)}\rangle$ is $\langle\Vec{r}|m\rangle$ at energy $E_m=E_g+\omega$(see Eq.\ref{eq-5}). The complex-valued on-shell contribution introduces an intermediate phase, which is partly responsible for the nonzero value of the minimum total cross section for the two-photon ATI process in the Fano lineshape \cite{PhysRevA.103.033103}. In contrast, the minimum total cross-section is normally expected to be zero for an autoionizing state that can decay into only one continuum.

\begin{figure}[htbp]
 \subfigure[] {\label{fig:a}\includegraphics[scale=0.5]{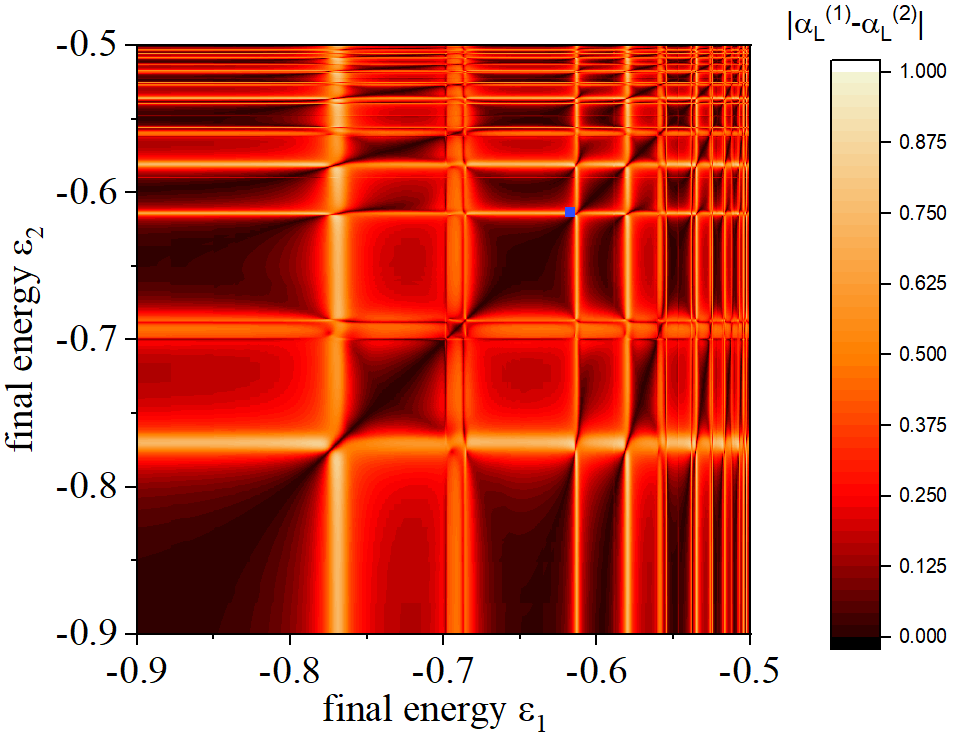}}
 \subfigure[] {\label{fig:b}\includegraphics[scale=0.29]{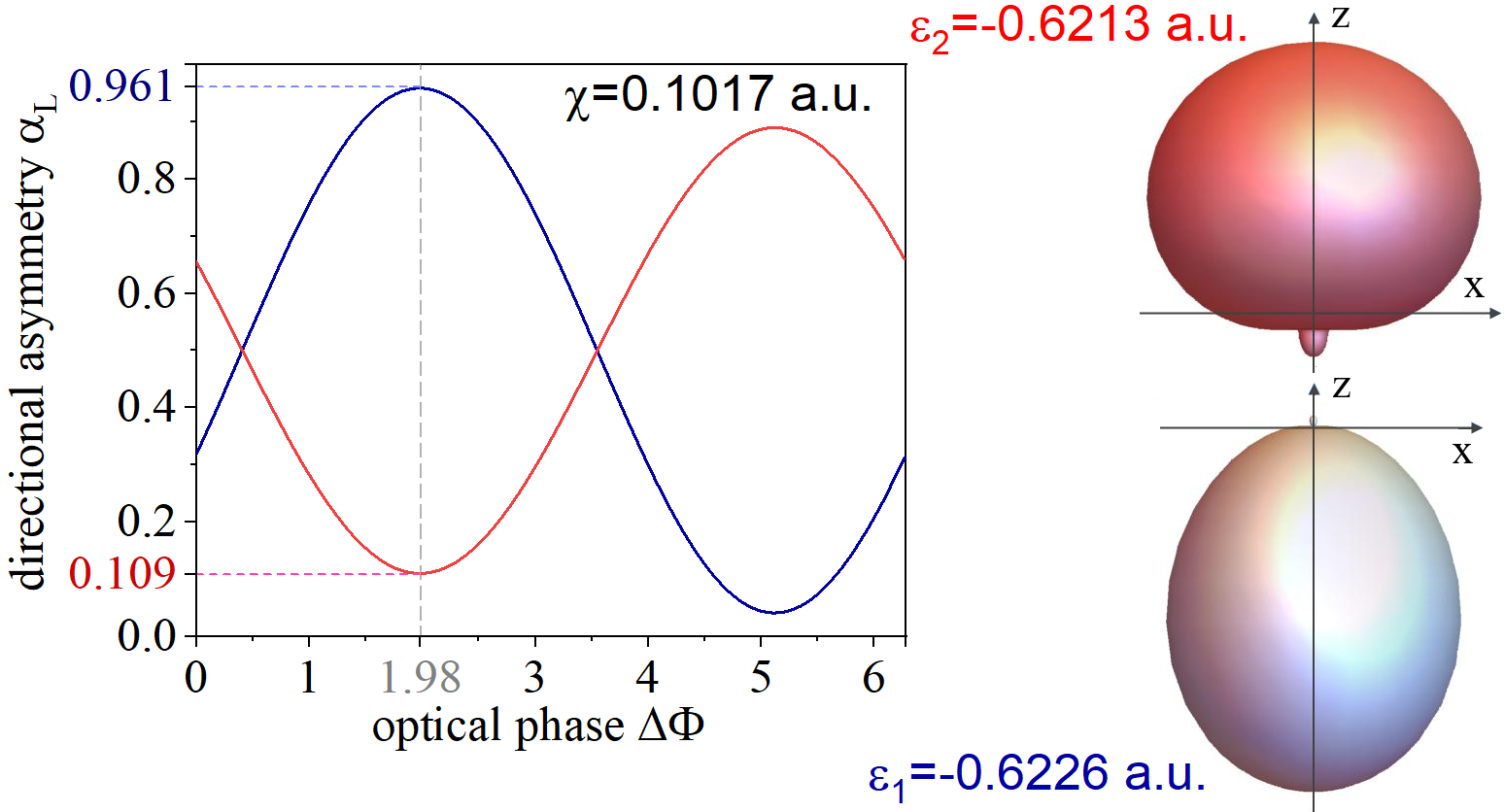}}
\hfill
\caption{(a). The differences in photoelectron directional asymmetry $|\alpha_L^{(1)}-\alpha_L^{(2)}|$ are shown for all possible ($\epsilon_1$,$\epsilon_2$) from -0.5 $a.u.$ to -0.9 $a.u.$. $\Delta\Phi$ and $\chi$ are chosen to optimize the difference $|\alpha_L^{(1)}-\alpha_L^{(2)}|$ at each value of the two energies on the plane. The bright grids indicate places where the energy pairs have a large angular asymmetry, and they show a close overlap with the ``resonance mesh" of horizontal and vertical stripes. The blue point near the S-wave $2p^2$ resonance indicates the energies considered in (b). (b). An example demonstrating how frequency-sensitive control can almost completely redirect the photoelectrons. The parameters are given in the figure. At $\Delta\Phi=0.63 \pi$(dashed vertical line) $\alpha_L^{(1)}-\alpha_L^{(2)}$ is maximized, with the values marked on the ticks. A polar plot of $\frac{dW(\theta)}{d\Omega}$ using the parameters indicated by the dashed lines is given in the right panel. With two very close frequencies, one that drives most photoelectrons along the polarization axis $+{\hat z}$,  while the other opposite to that direction. }
\label{3}
\end{figure}

Based on the discussions of $A(\chi,\epsilon)$ and $\varphi(\epsilon)$, we now explore the possibility of redirecting the photocurrent through frequency control. As demonstrated in Fig.~\ref{2},  the asymmetric phase $\varphi$ changes rapidly with energy when across the resonance, therefore with a fixed optical phase $\Delta\Phi$, a small change of $\epsilon$ can flip the escape direction of the photoelectrons between upper and lower halves of emission sphere. Now we consider the difference of $\alpha_L$ at two energies $\epsilon_1$ and $\epsilon_2$, using the same optical quantities ($\Delta\Phi$,$\chi$).  This gives an expression for $(\alpha_L^{(1)}-\alpha_L^{(2)})$, namely
\begin{equation}
\begin{split}
    \alpha_L^{(1)}-\alpha_L^{(2)}&=A_1\cos{(\Delta\Phi-\varphi_1)}-A_2\cos{(\Delta\Phi-\varphi_2)}\\
    &=Re\left[e^{i\Delta\Phi}(A_1e^{-i\varphi_1}-A_2e^{-i\varphi_2})\right]
\end{split}
\end{equation}
For this exploration $\Delta\Phi=-arg(A_1e^{-i\varphi_1}-A_2e^{-i\varphi_2})$, and $\chi$ is chosen to  maximize 
$|A_1e^{-i\varphi_1}-A_2e^{-i\varphi_2}|$ at each pair of energies. 
Next we scanned through all the ($\epsilon_1$,$\epsilon_2$) from -0.5 $a.u.$ to -0.9 $a.u.$ to search for candidates that have a large orientation difference. The maximum $|\alpha_L^{(1)}-\alpha_L^{(2)}|$ for all the energy points are shown in Fig.~\ref{3} (a).

The bright grids represent regions where the angular asymmetry difference $|\alpha_L^{(1)}-\alpha_L^{(2)}|$ is large, which generally tends to happen when at least one energy is close to a resonance. When both $\epsilon_1$,$\epsilon_2$ are far from resonances $|\alpha_L^{(1)}-\alpha_L^{(2)}|$ is small, which is expected since according to Fig.~\ref{2} in those regions $\varphi_{1,2}$ show little difference from each other. The points of greatest interest are near the intersections of the grids, where  $|\alpha_L^{(1)}-\alpha_L^{(2)}|$ rapidly changes, i.e. in those regions where both energies are on or near resonance. A specific example point has been selected near the $S$-wave $2p^2$ resonance, shown in Fig.~\ref{3}(a) as a blue point, for the following analysis.
%

The blue point is at $\epsilon_1=-0.6226$ $a.u.$ and $\epsilon_2=-0.6213$ $a.u.$, which go across the S-wave $2p^2$ resonance at -0.6222 $a.u.$ with width $\Gamma=2.36\times10^{-4}$ $a.u.$. The corresponding values of the other key parameters are  $\chi=0.1017$, $\Delta\Phi=0.63 \pi$. Their directional asymmetry parameters $\alpha_L^{(1),(2)}$ versus $\Delta\Phi$ are presented in Fig.~\ref{3}(b), and they are entirely out of phase from each other; when $\Delta\Phi=0.63 \pi$, both of the asymmetry parameters reach their corresponding extrema with values 0.961 and 0.109. The choice of $\chi$ maximizes this disparity, and it enables the one-photon transition(p-wave) to be strong enough to interfere with the strong on-resonance two-photon transition(s-wave). 
To examine the physics at those extremum points, the PAD ${dW(\theta)}/{d\Omega}$ is calculated with the parameters cited above, shown in Fig.~\ref{3}(b) right panel. This demonstrates how different the photoejection directions can be at $\epsilon_1=-0.6226$ $a.u.$($\omega=31.032$ $eV$) and $\epsilon_2=-0.6213$ $a.u.$ ($\omega=31.049$ $eV$). The central energies $\epsilon_1,\epsilon_2$ have been convolved over a resolution of $\pm3.5\times10^{-4}$ $a.u.$($0.01$ $eV$), in order to simulate a realistic experiment with finite resolution. 

\begin{figure}[htbp]
  \includegraphics[scale=0.45]{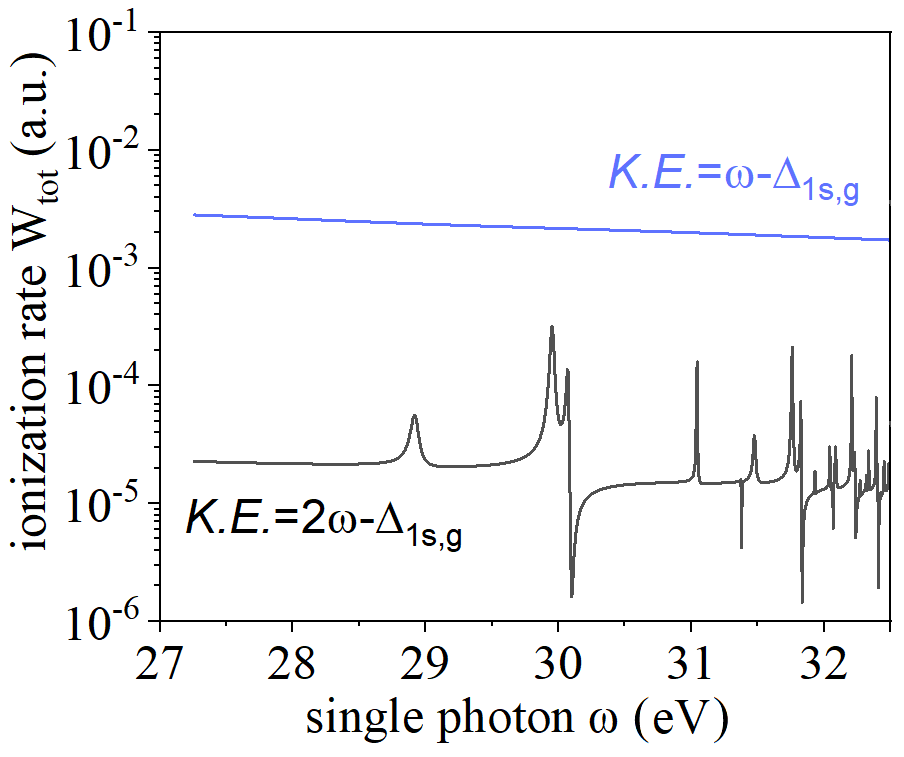}
    \caption{ The ionization rate of photoelectron escaping with different kinetic energies($K.E.$), with laser intensities ${I}_{\omega}=2.0\times10^{13} $ W cm$^{-2}$ and ${I}_{2\omega}=1.10\times10^{12} $ W cm$^{-2}$. $\Delta_{1s,g}=E_{1s}-E_g=24.58$ $eV$ is the energy difference from the $1s$ threshold and the ground state. The ionization rate at $\omega-\Delta_{1s,g}$ is about $100$ times larger than the rate for $2\omega-\Delta_{1s,g}$, which makes it challenging to detect the directional asymmetry properties of the faster electrons, although the two different energies are readily discriminated. }
  \label{4}
\end{figure}

One realization of the reduced field at $\chi=\mathcal{E}_{\omega}^2/{\mathcal{E}_{2\omega}}=0.1017$ $a.u.$ is to use lasers with intensities ${I}_{\omega}=2.0\times10^{13} $ W cm$^{-2}$(fundamental) and ${I}_{2\omega}=1.10\times10^{12} $ W cm$^{-2}$(second harmonic), 
 where $\mathcal{E}_{\nu}=\sqrt{{2{I}_{\nu}}/{\epsilon_0 c}} \ \text{a.u.}/{5.1422\times10^{9} }\text{V cm}^{-1}$. Based on these laser intensities, we analyze here the reasonableness of a possible implementation of this frequency-sensitive control scheme. 
The total rates for asymmetric photoejection in Fig.~\ref{3}(b) are, $W_{\rm tot}(\epsilon_1)=2.54\times10^{-5}$ $a.u.$ and $W_{\rm tot}(\epsilon_2)=3.08\times10^{-5}$ $a.u.$, but these only include photoelectrons that escaped after absorbing $2\omega$ of energy.
There are extra photoelectrons that escape from the two-photon pathway intermediate process, i.e. from absorbing a single photon of frequency $\omega$, which have not been discussed, because they are not being controlled by the optical interference effect. The ionization rate for the ``intermediate state leaked" $\omega$-absorption process is much stronger, usually by a factor of $100$, than the $2\omega$-absorption process. At energies $\epsilon_1$ and $\epsilon_2$, their corresponding intermediate ionization rates are around $1.99\times10^{-3}$ $a.u.$. Thus to observe the experimental interference control predicted in the present study, electron energy discrimination is required.
The ionization rates $W_{\rm tot}$ for both the processes covering final state energy $\epsilon$ from $-0.9$ to $-0.5$ $a.u.$ are given in Fig.~\ref{4}. The dominance of intermediate ionization is rather typical for most ATI processes, except for a few cases when the intermediate states hit a near-zero ionization minimum, which are found in some alkaline earth atoms such as Ca, Sr, and Ba. Searching for proper ionization processes that can suppress the intermediate leakage can be a goal for our future study. 
For helium where the intermediate states lie in a flat continuum for the energy range considered in this study, the ratio between for $2\omega$- and $\omega$-absorption is around $\mathcal{E}_{\omega}^2$ $a.u.$-- very tiny for electric field below the tunneling region, which implies that the optical control scheme is not highly efficient.

\section{Conclusion}

To conclude, the present treatment of energy-dependent coherent control over the directional asymmetry of helium ionization has identified the optical phase difference $\Delta\Phi$ and reduced electric field strength $\chi$ that largely enhance the degree of control of the directional photoejection asymmetry.  Our study suggests an alternative way of redirecting the photoelectrons by changing the laser frequency but with a fixed relative phase and field strength ratio, and we presented an example using this frequency-sensitive controlling scheme to redirect photoelectron with final state energies across the S-wave $2p^2$ resonance. However, due to the existence of intermediate-state photoionization, the coherent control can only influence a small fraction of the total electron current which makes any experimental test of our predictions  demanding. Future studies of the alkaline earth atoms that have a continuum ionization minimum of either the Fano- or Cooper-type might circumvent this issue.

\begin{acknowledgments}
This work was supported by the U.S. Department of Energy, Office of Science, Basic Energy Sciences, under Award No. DE-SC0010545.
\end{acknowledgments}

\nocite{*}

\bibliographystyle{unsrt}
\bibliography{resonance_control}

\end{document}